\documentstyle[amssymb,aps,prb,twocolumn,psfig]{revtex}

\begin{document}

%
\twocolumn[\hsize\textwidth\columnwidth\hsize\csname@twocolumnfalse\endcsname 
\title {\ Transition Radiation of Moving Abrikosov Vortices. }
\author{ O.V. Dolgov}
\address{Max- Planck-Institut f\"{u}r Festk\"{o}rperforschung, \\
Heisenbergstrasse 1,
D-70569 Stuttgart, Germany}
\author{ N. Schopohl}
\address{Eberhard-Karls-Universit\"{a}t T\"{u}bingen,
Institut f\"{u}r Theoretische Physik, \\
Auf der Morgenstelle 14, \\
D-72076 T\"{u}bingen, Germany}

\maketitle
\begin{abstract}
We predict that Abrikosov vortices moving towards the surface of a
superconductor emit electromagnetic radiation into free space.
Passage of a single vortex line travelling at constant speed ${\bf v}$
from a superconducting half space, $x<0$, towards the vacuum half space, $x>0$,  
is discussed in detail. The frequency distribution of
the radiated intensity displays a pronounced maximum at
microwave frequencies around $\omega \simeq |v_{x}|/\lambda $, where $\lambda $
is the magnetic penetration depth. Coherent motion of a
lattice of flux lines leads to constructive interference and increases the strength of the radiated power per
vortex line by a large factor. Also the radiation source may be used to measure the
velocity of Abrikosov vortices and to determine the magnetic penetration depth. 
\end{abstract}
\bigskip
]%
%

A charged particle, moving inside a homogenous medium at constant speed $%
{\bf v}$ and not exceeding the phase velocity $c$ of light in the medium,
cannot radiate electromagnetic waves. However, when the particle passes from
one medium into another with different electromagnetic properties there may
occur transition radiation\cite{frank},\cite{Ginzburg}. In sharp contrast to
Cherenkov radiation there exists no lower bound on the velocity of a charged
particle for transition radiation.

In the following we suggest that the phenomenon of transition radiation
should also occur for electromagnetic objects like Abrikosov vortices moving
towards the surface of a superconductor. In our calculation we assume all
vortex lines being orientated parallel to the $\widehat{{\bf z}}$-axis and
we assume $\widehat{{\bf z}}\cdot {\bf v}=0$.

Let us first consider a single Abrikosov vortex at rest\cite{Abrikosov}\cite
{Shmidt}. Such a vortex is a topological object inside an ordered medium,
namely a line defect in a superconductor. Also it is a quantum object. The
gradient of the phase of the pairing wave function drives Cooper pairs (
i.e. pairs of correlated electrons near the Fermi surface carrying a small
center of mass momentum) in circles around the vortex axis, while the
modulus of the pairing potential vanishes at the center of the vortex core
(say the node is at position ${\bf R}$ in the $xy$-plane). The flow of mass
around the vortex axis displays a {\em quantized }circulation in units of $%
\frac{h}{2m}$, and the magnetic flux associated with such a line defect is
quantized in units of $\Phi _{0}=\frac{hc}{2e}$, the quantum of flux. Due to
Meissner screening the magnetic field around the vortex decays exponentially
fast on the scale of the London penetration depth $\lambda $ with increasing
distance $\left| {\bf r-R}\right| \gg \lambda $ to the vortex core: 
\begin{equation}
{\bf B}({\bf r})=\widehat{{\bf z}}\frac{\Phi _{0}}{2\pi \lambda ^{2}}\sqrt{%
\frac{\lambda }{\left| {\bf r-R}\right| }}\exp (-\frac{\left| {\bf r-R}%
\right| }{\lambda })
\end{equation}
In sharp contrast to this fast decay of the magnetic field ${\bf B}({\bf r})$
is the asymptotic behaviour of the vector potential ${\bf A}({\bf r})$
around an Abrikosov vortex, which is connected to the magnetic field via $%
{\bf B}({\bf r})=%
\mathop{\rm rot}%
{\bf A}({\bf r})$: 
\begin{equation}
\begin{array}{ccc}
{\bf A}({\bf r})=\frac{\Phi _{0}}{2\pi }\,\widehat{{\bf z}}\wedge \frac{{\bf %
r-R}}{\left| {\bf r-R}\right| ^{2}} & \text{for} & \left| {\bf r-R}\right|
\gg \lambda
\end{array}
\end{equation}
Note that ${\bf A}({\bf r})$ is independent on $\lambda $ far outside the
London core (apart from exponentially small correction terms).

This slow decay of the vector potential around a single Abrikosov vortex is
reminiscent to the behaviour of the vector potential in the exterior region
of a long solenoid, like in an {\em Aharonov-Bohm} experiment. Consider a
solenoid consisting out of a piece of infinitesimally thin wire wound
helically around the axis $\widehat{{\bf z}}$ of an infinitely long cylinder
with radius $r_{\sigma }$. Through the wire of the solenoid flows a suitable
electrical current $I_{\sigma }$ to generate a constant magnetic field ${\bf %
B}^{\left( \sigma \right) }=B^{\left( \sigma \right) }\widehat{{\bf z}}$ 
{\em inside}, but {\em outside} the solenoid the magnetic field is zero. The
vector potential ${\bf A}^{\left( \sigma \right) }$ generated by the
solenoid in its exterior region is the gradient of a 'phase': 
\begin{equation}
{\bf A}^{\left( \sigma \right) }({\bf r})=\left\{ 
\begin{array}{c}
\frac{B^{\left( \sigma \right) }}{2}\widehat{{\bf z}}\wedge \left( {\bf r}-%
{\bf R}\right) \text{\hspace{0.15in}for\hspace{0.15in}}\left| {\bf r}-{\bf R}%
\right| \leq r_{\sigma } \\ 
\frac{\Phi _{\sigma }}{2\pi }\,{\bf \nabla }\vartheta \left( {\bf r}-{\bf R}%
\right) \text{\hspace{0.1in}for\hspace{0.1in}}\left| {\bf r}-{\bf R}\right|
>r_{\sigma }
\end{array}
\right.
\end{equation}
where $e^{i\vartheta \left( {\bf r}\right) }=\frac{x+iy}{\sqrt{x^{2}+y^{2}}}$%
, ${\bf \nabla }\vartheta ({\bf r})=\widehat{{\bf z}}\wedge \frac{{\bf r}}{%
\left| {\bf r}\right| ^{2}}$ and ${\bf r}=(x,y)$. Consider a closed path $%
\Gamma $ in the exterior region of the solenoid with winding number $1$
around the $\widehat{{\bf z}}-$axis. Integration of ${\bf A}^{\left( \sigma
\right) }({\bf r})$ along $\Gamma $ leads to: 
\begin{equation}
\oint_{\Gamma }d{\bf r\cdot A}^{\left( \sigma \right) }({\bf r})=\Phi
_{\sigma }=\int_{\Omega }d^{2}r\ \widehat{{\bf z}}\cdot 
\mathop{\rm rot}%
{\bf A}^{\left( \sigma \right) }({\bf r})=\pi r_{\sigma }^{2}\,B^{\left(
\sigma \right) }
\end{equation}
(apply Stokes theorem to the area $\Omega $ enclosed by $\Gamma $) . So, the
constant $\Phi _{\sigma }$ is equal to the flux passing through the
solenoid. Adjusting the control current $I_{\sigma }$ such that $\Phi
_{\sigma }$ becomes equal to one quantum of flux, $\left| \Phi _{\sigma
}\right| =\Phi _{0}$ , and taking the limit $r_{\sigma }\rightarrow 0^{+}$,
\ the thin solenoid produces (in its rest frame) a singular electromagnetic
field at its center: 
\begin{eqnarray}
{\bf B}^{\left( \sigma \right) }({\bf r}) &=&\ \Phi _{\sigma }\widehat{{\bf z%
}}\,\delta ^{(2)}({\bf r}-{\bf R})  \nonumber \\
{\bf E}^{\left( \sigma \right) }({\bf r}) &=&{\bf 0}
\end{eqnarray}
Consider now a thin solenoid travelling relative to the lab frame at a
constant speed ${\bf v}$ along the trajectory ${\bf R}={\bf v}t$. Such a
solenoid is surrounded by a {\em time dependent} Aharonov-Bohm type vector
potential giving rise not only to a singular magnetic field ${\bf B}^{\left(
\sigma \right) }({\bf r},t)$ but also to a singular electric field ${\bf E}%
^{\left( \sigma \right) }({\bf r},t)$: 
\begin{eqnarray}
{\bf B}^{\left( \sigma \right) }({\bf r,}t) &=&\Phi _{\sigma }\widehat{{\bf z%
}}\,\delta ^{(2)}({\bf r}-{\bf v}t) \\
{\bf E}^{\left( \sigma \right) }({\bf r,}t) &=&-\frac{{\bf v}}{c}\wedge {\bf %
B}^{\left( \sigma \right) }({\bf r,}t)  \nonumber
\end{eqnarray}
It is consistent (and convenient) to associate with the moving thin solenoid
a ${\em conserved}$ charge-current density via: 
\begin{eqnarray}
4\pi \rho ^{(\sigma )}({\bf r,}t) &=&%
\mathop{\rm div}%
{\bf E}^{\left( \sigma \right) }({\bf r,}t)  \label{external sources} \\
\frac{4\pi }{c}{\bf J}^{(\sigma )}({\bf r,}t) &=&%
\mathop{\rm rot}%
{\bf B}^{\left( \sigma \right) }({\bf r,}t)-\frac{1}{c}\partial _{t}{\bf E}%
^{\left( \sigma \right) }({\bf r,}t)  \nonumber
\end{eqnarray}
Since in the geometry under consideration translational invariance is broken
at $x=0$ it is natural to require that no current must flow from the region $%
x<0$ (superconductor) into the region $x>0$ (vacuum): 
\begin{equation}
\lim_{x\rightarrow 0}\widehat{{\bf n}}\cdot {\bf J}^{(\sigma )}({\bf r,}t)=0
\label{boundary condition source term}
\end{equation}
($\widehat{{\bf n}}$ denotes the normal unit vector of the interface). By a
simple trick we are able to fulfill this boundary condition (which actually
avoids solving an integral equation problem of the Wiener-Hopf type). Let us
superpose to the field around a single flux line situated at its
(instantaneous) position ${\bf R}={\bf v}t$ another singular field generated
by a mirror vortex (with {\em opposite} magnetic moment) situated at the 
{\em mirror reflected} position $\widetilde{{\bf R}}=\widetilde{{\bf v}}t$ ,
where $\widetilde{{\bf v}}=(-v_{x},v_{y})$. The resulting fields (inside the
region $x<0$) are then: 
\begin{eqnarray}
{\bf B}^{\left( \sigma \right) }({\bf r},t) &=&\ \Phi _{\sigma }\widehat{%
{\bf z}}\,\left[ \delta ^{(2)}({\bf r}-{\bf v}t)-\delta ^{(2)}({\bf r}-%
\widetilde{{\bf v}}t)\right]  \label{self fields} \\
{\bf E}^{\left( \sigma \right) }({\bf r},t) &=&\Phi _{\sigma }\widehat{{\bf z%
}}\wedge \left[ \frac{{\bf v}}{c}\delta ^{(2)}({\bf r}-{\bf v}t)-\frac{%
\widetilde{{\bf v}}}{c}\delta ^{(2)}({\bf r}-\widetilde{{\bf v}}t)\right] 
\nonumber
\end{eqnarray}
Inserting these expressions for ${\bf B}^{\left( \sigma \right) }$ and ${\bf %
E}^{\left( \sigma \right) }$ into Eq.(\ref{external sources}) there results\
a {\em conserved} charge-current distribution which generates the
asymptotics of the Aharonov-Bohm type vector potential around a {\em moving}
Abrikosov vortex line far outside the London core region. By construction
this charge-current distribution also fulfills the boundary value condition
Eq.(\ref{boundary condition source term}).

Let us next discuss the effect of Meissner screening. Since a superconductor
is an {\em ordered} medium it will react to a specified external influence
by a specific response{\em \ }that takes into account the formation of
induced currents and of induced polarization charges inside the sample. This
means a {\em parametrically} driven flux line moving at speed ${\bf v}$
inside a medium with Meissner screening (imagine an external agent who pulls
the solenoid along) is surrounded by {\em total} current and charge
densities of the form 
\begin{eqnarray}
{\bf J}^{\left( tot\right) }({\bf r},\omega ) &=&{\bf J}^{(ind)}({\bf r}%
,\omega )+{\bf J}^{(\sigma )}({\bf r},\omega )  \label{charge-current} \\
\rho ^{\left( tot\right) }({\bf r},\omega ) &=&\rho ^{(ind)}({\bf r},\omega
)+\rho ^{(\sigma )}({\bf r},\omega )  \nonumber
\end{eqnarray}
Here ${\bf J}^{(ind)}$ and $\rho ^{(ind)}$ represent the induced response of
the ordered medium to the applied drive. Explicit expressions for the drive
terms ${\bf J}^{(\sigma )}({\bf r},\omega )$ and $\rho ^{(\sigma )}({\bf r}%
,\omega )$ follow inserting Eq.(\ref{self fields}) into Eq.(\ref{external
sources}) and calculating the Fourier transforms with respect to time. Since
the induced response is retarded in time it is convenient to switch to the
frequency representation. Suitable expressions for $\rho ^{(ind)}({\bf r}%
,\omega )$ and ${\bf J}^{(ind)}({\bf r},\omega )$ follow from standard
linear response theory. Since any {\em \ local} perturbation inside the bulk
of a superconductor heals typically on the length scale of the coherence
length $\xi $ of the Cooper pairs ($\xi \ll \lambda $ in extreme type-II
materials), and since the phenomenon we study is related to low frequencies
in the microwave regime, a suitable (local) expression for the induced
charge-current distribution is: 
\begin{eqnarray}
{\bf J}^{(ind)}({\bf r},\omega ) &=&-\frac{c}{4\pi }\ \frac{c}{i\omega }%
\frac{1}{\lambda _{\omega }^{2}}\ {\bf E}^{\left( tot\right) }({\bf r}%
,\omega )  \label{J-ind} \\
\rho ^{(ind)}({\bf r},\omega ) &=&\frac{1}{i\omega }%
\mathop{\rm div}%
{\bf J}^{(ind)}({\bf r},\omega )  \nonumber
\end{eqnarray}
In a '{\em two fluid' } type description of the electromagnetic response of
a superconductor it is found that the length $\lambda _{\omega }$ is a
frequency dependent (but local) penetration depth, that interpolates between
the skin penetration depth, which is characteristic for the normal metallic
state at frequencies above $2\Delta $, to the magnetic penetration depth
(MPD) at zero frequency: 
\begin{equation}
\frac{1}{\lambda _{\omega }^{2}}=\frac{\omega _{pl}^{2}}{c^{2}}\frac{\omega 
}{\omega +i\gamma }\Theta (\left| \omega \right| -2\Delta )+\frac{1}{\lambda
^{2}}\Theta (2\Delta -\left| \omega \right| )
\end{equation}
Here $\omega _{pl}=\sqrt{\frac{4\pi n\,e^{2}}{m^{\ast }}}$ is the bulk
plasma frequency, $\gamma $ describes quasiparticle damping, $\lambda
=\lambda \left( \gamma \right) $ denotes MPD of a superconductor \cite{Nam}.
In the clean limit (and at zero temperature) lim$_{\gamma \rightarrow
0}\lambda \left( \gamma \right) =\frac{c}{\omega _{pl}}=$ $\lambda _{L}$,
the standard definition of the London penetration depth. Of course, the
nonlocality of the electromagnetic response kernel becomes eventually
important very near to the surface (e.g. anomalous skin effect).

In the presence of Meissner screening the (total) electromagnetic field
around a travelling thin solenoid consists out of {\em two} parts. The {\em %
first} part is entirely due to the Meissner screening currents which flow in
the exterior region of the solenoid. This part actually represents (within
the accuracy of local London electrodynamics) the electromagnetic fields
around a moving Abrikosov vortex, provided the flux inside the solenoid is
fixed with the opposite sign: $\Phi _{\sigma }=-\Phi _{0}$. The {\em second}
part originates from the infinitesimal small {\em interior} region of the
thin solenoid, where there exists (in the rest frame of the solenoid) an
artificially fixed singular magnetic field: $B^{\left( \sigma \right) }=$ $%
\frac{\Phi _{\sigma }}{\pi r_{\sigma }^{2}}$. This feature of the thin
solenoid model is in sharp contrast to what really happens at the center of
an Abrikosov vortex, where the pairing wave function displays a node and
where the magnetic field assumes a finite and (almost) constant value inside
the inner core of the vortex , a region of size $\xi \ll \lambda $. As a
result, the self fields ${\bf E}^{\left( \sigma \right) }({\bf r},\omega )$
and ${\bf B}^{\left( \sigma \right) }({\bf r},\omega )$, which originate
from the {\em interior} region of the solenoid, should be subtracted from
the total fields ${\bf E}^{\left( tot\right) }({\bf r},\omega )$ and ${\bf B}%
^{\left( tot\right) }({\bf r},\omega )$ \ in order to determine the physical
fields ${\bf E}({\bf r},\omega )$ and ${\bf B}({\bf r},\omega )$ around an
Abrikosov vortex line: 
\begin{eqnarray}
{\bf B}^{\left( tot\right) }({\bf r},\omega ) &=&\ {\bf B}({\bf r},\omega )+%
{\bf B}^{\left( \sigma \right) }({\bf r},\omega )  \label{EB-tot} \\
{\bf E}^{\left( tot\right) }({\bf r},\omega ) &=&\ {\bf E}({\bf r},\omega )+%
{\bf E}^{\left( \sigma \right) }({\bf r},\omega )  \nonumber
\end{eqnarray}
The fields ${\bf E}^{\left( tot\right) }({\bf r},\omega )$ and ${\bf B}%
^{\left( tot\right) }({\bf r},\omega )$ are solutions to the Maxwell
equations in the presence of the total charge-current distribution specified
in Eq.(\ref{charge-current}): 
\begin{eqnarray}
&&{\bf \;}\frac{4\pi }{c}\left[ {\bf J}^{\left( \sigma \right) }({\bf r}%
,\omega )+{\bf J}^{\left( ind\right) }({\bf r},\omega )\right]  \\
&=&{\bf 
\mathop{\rm rot}%
B}^{\left( tot\right) }({\bf r},\omega )+\frac{i\omega }{c}{\bf E}^{\left(
tot\right) }({\bf r},\omega )  \nonumber
\end{eqnarray}
Upon insertion of Eq.(\ref{J-ind}) this transforms into: 
\begin{eqnarray}
&&{\bf \;}\frac{4\pi }{c}{\bf J}^{\left( \sigma \right) }({\bf r},\omega )-\ 
\frac{c}{i\omega }\frac{1}{\lambda _{\omega }^{2}}\ \left[ {\bf E}({\bf r}%
,\omega )+{\bf E}^{\left( \sigma \right) }({\bf r},\omega )\right]  \\
&=&{\bf 
\mathop{\rm rot}%
}\left[ {\bf B}({\bf r},\omega )+{\bf B}^{\left( \sigma \right) }({\bf r}%
,\omega )\right] +\frac{i\omega }{c}\left[ {\bf E}({\bf r},\omega )+{\bf E}%
^{\left( \sigma \right) }({\bf r},\omega )\right]   \nonumber
\end{eqnarray}
Next we insert ${\bf J}^{\left( \sigma \right) }$ from the definition Eq.(%
\ref{external sources}) and obtain: 
\begin{equation}
{\bf \;}\ -\frac{c}{i\omega }\frac{1}{\lambda _{\omega }^{2}}\ \left[ {\bf E}%
({\bf r},\omega )+{\bf E}^{\left( \sigma \right) }({\bf r},\omega )\right] =%
{\bf 
\mathop{\rm rot}%
B}({\bf r},\omega )+\frac{i\omega }{c}{\bf E}({\bf r},\omega )
\end{equation}
Using the Faraday law 
\[
{\bf 
\mathop{\rm rot}%
\,}{\bf E(r,}\omega )-\frac{i\omega }{c}{\bf B}({\bf r},\omega )={\bf 0}
\]
it is not difficult to derive {\em dynamical} London equations for the
physical electromagnetic fields around a moving vortex line: 
\begin{eqnarray}
{\bf 
\mathop{\rm rot}%
\mathop{\rm rot}%
\,}{\bf E(r,}\omega {\bf )}+\left( \frac{1}{\lambda _{\omega }^{2}}-\frac{%
\omega ^{2}}{c^{2}}\right) {\bf E(r,}\omega {\bf )} &=&{\bf \;}\frac{-1}{%
\lambda _{\omega }^{2}}\,{\bf E}^{\left( \sigma \right) }{\bf (r,}\omega 
{\bf )}  \label{dynamical London equations} \\
\left[ {\bf -\Delta }+\left( \frac{1}{\lambda _{\omega }^{2}}-\frac{\omega
^{2}}{c^{2}}\right) \right] {\bf B(r,}\omega {\bf )} &=&{\bf \;}\frac{-1}{%
\lambda _{\omega }^{2}}\,{\bf B}^{\left( \sigma \right) }{\bf (r,}\omega 
{\bf )}  \nonumber
\end{eqnarray}
For $x>0$ we have $\frac{1}{\lambda _{\omega }^{2}}\equiv 0$, of course,
since there is nothing in vacuum. In the {\em static} limit, $\left| {\bf v}%
\right| \rightarrow 0$ and $\omega \rightarrow 0$, the second equation
reduces to the standard London equation determining the magnetic field
around an Abrikosov vortex line at rest\cite{de Gennes},\cite{Abrikosov}, 
\cite{Shmidt}. Note that for simplicity we assume here that there exists no
external static magnetic field in vacuum, i.e. $\lim_{\omega \rightarrow 0}%
{\bf B(r,}\omega )={\bf 0}$ for $x>0$.

Next we discuss scattering solutions to our dynamical London equations,
which describe the transition radiation effect for moving Abrikosov
vortices. Before we expand details of the new radiation source let us
comment on an interesting difference to standard transition radiation\cite
{frank},\cite{Ginzburg}. In the case of standard transition radiation the
moving particle, say an electron or a proton, actually exists on either side
of the interface passing from one medium into another with different
electromagnetic properties. In the case of a moving vortex the 'particle'
(i.e.the travelling singular line representing the center of the vortex),
exists only inside the superconductor, but ceases to exist on the other side
of the interface (in vacuum).

A Fourier transformation of the dynamical London equations with respect to
the $y-$coordinate simplifies the problem and leads to coupled ordinary
differential equations. It is sufficient to consider the tangential
component $E_{y}(x)=E_{y}(x,k_{y},\omega )$ of the electric field, since the
normal component $E_{x}(x)=E_{x}(x,k_{y},\omega )$ follows from knowledge of 
$\partial _{x}E_{y}(x)$. From both components of the electric field the
magnetic field $B_{z}(x)=B_{z}(x,k_{y},\omega )$ may be calculated using
Faraday's law. Guided by the structure of our problem we make now the
following ansatz for $E_{y}(x)$ in terms of three unknown coefficients $%
a_{S} $ , $b_{S}$ and $a_{V}$ , and in terms of three (complex) wave numbers 
$q_{\lambda }$ , $q_{\infty }$ and $K_{\omega }$: 
\begin{eqnarray}
&& \\
E_{y}(x) &=&\left\{ 
\begin{array}{ccc}
a_{S}\exp \left( q_{\lambda }x\right) +b_{S}\left( e^{iK_{\omega
}x}+e^{-iK_{\omega }x}\right) & \text{for} & x<0 \\ 
a_{V}\exp \left( -q_{\infty }x\right) & \text{for} & x>0
\end{array}
\right.  \nonumber
\end{eqnarray}
\begin{eqnarray}
\;q_{\infty } &=&\sqrt{k_{y}^{2}{\bf \ -}\frac{\omega ^{2}}{c^{2}}}%
\;,\;q_{\lambda }=\sqrt{k_{y}^{2}{\bf \ +}\frac{1}{\lambda _{\omega }^{2}}\ 
{\bf -}\frac{\omega ^{2}}{c^{2}}}\;  \nonumber \\
\;K_{\omega } &=&\frac{\omega -k_{y}v_{y}}{v_{x}}\;
\end{eqnarray}
At infinity ($\left| x\right| \rightarrow \infty $) we pose the usual
radiation boundary conditions. So, in order to avoid exponential growth for
\ $x\rightarrow \pm \infty $ , the branch cut of the square root is defined
such that $%
\mathop{\rm Re}%
q_{\lambda }>0$ and $%
\mathop{\rm Re}%
q_{\infty }>0$. The coefficient $b_{S}$ follows from the requirement that $%
E_{y}(x)$ should be a solution to the differential equation in the half
space $x<0$. The coefficients $a_{S}$ and $a_{V}$ \ follow from the standard
boundary value conditions for the tangential and normal components of the
electric field at $x=0$. The results are: 
\begin{eqnarray}
b_{S} &=&\frac{%
\mathop{\rm sign}%
(v_{x})}{c}\,\frac{\Phi _{0}}{\lambda _{\omega }^{2}}\,\frac{\left( \frac{%
v_{x}}{c}q_{\lambda }\right) ^{2}-\frac{v_{y}}{c}\,k_{y}\,K_{\omega }}{\frac{%
1}{\lambda _{\omega }^{2}}-\frac{\omega ^{2}}{c^{2}}}\frac{1}{K_{\omega
}^{2}+\left| \frac{v_{x}}{c}\right| ^{2}q_{\lambda }^{2}}  \nonumber \\
a_{S} &=&\frac{2b_{S}\,\frac{\omega ^{2}}{c^{2}}\,q_{\lambda }}{\left( -%
\frac{\omega ^{2}}{c^{2}}\right) q_{\lambda }+\left( \frac{1}{\lambda
_{\omega }^{2}}-\frac{\omega ^{2}}{c^{2}}\right) q_{\infty }}  \nonumber \\
a_{V} &=&\frac{2b_{S}\,\left( \frac{1}{\lambda _{\omega }^{2}}-\frac{\omega
^{2}}{c^{2}}\right) \,q_{\infty }}{\left( -\frac{\omega ^{2}}{c^{2}}\right)
q_{\lambda }+\left( \frac{1}{\lambda _{\omega }^{2}}-\frac{\omega ^{2}}{c^{2}%
}\right) q_{\infty }}
\end{eqnarray}
The coefficient $b_{S}$ displays a resonance around $\omega =\omega _{pl}$,
while the coefficients $a_{S}$ and $a_{V}$ display also a resonance around $%
\omega =\frac{\omega _{pl}}{\sqrt{2}}$, the surface plasma frequency. The
solution for $E_{y}(x)$ in vacuum ($x>0$) describes an outward travelling
wave front, representing electromagnetic radiation. The radiation part is
entirely determined by oscillating terms, which exist only in the range of
wave numbers $\left| k_{y}\right| {\bf \ <\ }\left| \frac{\omega }{c}\right| 
$ , so that $q_{\infty }=-i%
\mathop{\rm sign}%
(\omega )\sqrt{\frac{\omega ^{2}}{c^{2}}-k_{y}^{2}}$ is purely imaginary.
For $\left| k_{y}\right| {\bf \ >\ }\left| \frac{\omega }{c}\right| $ the
modulus of the electromagnetic field in vacuum is exponentially small for $%
x\ >>\frac{c}{\left| \omega \right| }$ , and may be safely neglected.

Consider now a planar detector screen at a distance $x_{T}\gg \frac{c}{%
\left| \omega \right| }$ in the far zone in vacuum, with the plane of the
screen orientated parallel to the superconductor-vacuum interface (at $x=0$%
). The rate{\em \ }at which the emitted radiation is incident onto a target
area element $\ \widehat{{\bf n}}$ $da$ $=-\widehat{{\bf x}}\ da$ at
position ${\bf r}_{T}=(x_{T},y_{T})$ on the screen may be calculated from
the associated Poynting flux $\frac{c}{4\pi }\ \left[ {\bf E}\left( {\bf r}%
_{T},t\right) \wedge {\bf B}\left( {\bf r}_{T},t\right) \right] \cdot \left(
-\widehat{{\bf x}}\right) $. Let us denote the amount of energy $%
d\varepsilon ({\bf r}_{T})$ so radiated (per length unit $L$ of the vortex
line) onto an infinitesimal target area element $da_{T}=L\cdot dy_{T}$ by $%
\frac{d\varepsilon ({\bf r}_{T})}{da_{T}}$. Then the total energy $%
\varepsilon _{rad}$ radiated into free space (per unit length $L$ of the
vortex line) is obtained integrating $\frac{d\varepsilon ({\bf r}_{T})}{%
da_{T}}$ over time and over the the entire screen area (wide detector): 
\[
\frac{\varepsilon _{rad}}{L}=\frac{c}{4\pi }\int_{-\infty }^{\infty }\frac{%
d\omega }{2\pi }\int_{-\left| \frac{\omega }{c}\right| }^{\left| \frac{%
\omega }{c}\right| }\frac{dk_{y}}{2\pi }\ E_{y}{\bf (}x_{T},k_{y}{\bf ,}%
\omega {\bf )}B_{z}{\bf (}x_{T},-k_{y}{\bf ,-}\omega {\bf )} 
\]
Let us introduce scaled variables: $\overline{\omega }=\frac{\omega }{\omega
_{pl}}\,$, $\overline{\lambda }=\frac{\lambda \left( \gamma \right) }{%
\lambda _{L}}$ , $\,\overline{k}_{y}=\lambda _{L}k_{y}\;$,$\,\overline{q}%
_{\infty }=\lambda _{L}q_{\infty }\,$,$\,\overline{q}_{\lambda }=\lambda
_{L}q_{\lambda }$ and $\overline{v}_{x}=\frac{v_{x}}{c}$,$\;\overline{v}_{y}=%
\frac{v_{y}}{c}$ , so that energy is measured in units of $\hbar \omega
_{pl} $ , length is measured in units of $\lambda _{L}=\frac{c}{\omega _{pl}}
$ and velocity is measured in units of $c$. We may then rewrite the vacuum
amplitude $a_{V}=a\left( k_{y}{\bf ,}\omega \right) =\frac{\Phi _{0}}{c}%
\,A_{V}\left( \overline{k}_{y}{\bf ,}\overline{\omega }\right) $ in
dimensionless form: 
\begin{equation}
A_{V}=\frac{2}{\overline{\lambda }_{\omega }^{2}}\frac{%
\mathop{\rm sign}%
(v_{x})}{\left( \overline{\omega }-\overline{v}_{y}\overline{k}_{y}\right)
^{2}+\overline{v}_{x}^{2}\overline{q}_{\lambda }^{2}}\frac{\overline{q}%
_{\infty }\left( \overline{v}_{x}^{2}\overline{q}_{\lambda }^{2}+\overline{v}%
_{y}^{2}\overline{k}_{y}^{2}-\overline{v}_{y}\overline{k}_{y}\overline{%
\omega }\right) }{\left( \frac{1}{\overline{\lambda }_{\omega }^{2}}-%
\overline{\omega }^{2}\right) \overline{q}_{\infty }-\overline{\omega }^{2}%
\overline{q}_{\lambda }}
\end{equation}
It follows 
\begin{equation}
\frac{\varepsilon _{rad}}{L}=\left( \frac{\Phi _{0}}{2\pi \lambda _{L}}%
\right) ^{2}\,\frac{1}{2}\int_{-\frac{\pi }{2}}^{\frac{\pi }{2}}\frac{%
d\theta }{\pi }\int_{0}^{\infty }d\overline{\omega }\,\,\overline{\omega }%
\,\,\,\left| A_{V}\left( \overline{\omega }\sin \theta {\bf ,}\overline{%
\omega }\right) \right| ^{2}
\end{equation}
The function $I\left( \theta ,\overline{\omega }\right) =\overline{\omega }%
\,\,\cdot \,\left| A_{V}\left( \overline{\omega }\sin \theta {\bf ,}%
\overline{\omega }\right) \right| ^{2}$ \ is the angle-frequency
distribution of transition radiation emitted into free space by a single
moving Abrikosov vortex. Apparently, the dependence of $I\left( \theta ,%
\overline{\omega }\right) $ on the angle $\theta $ for a non relativistic
speed is such that it is practically constant for all directions, with the
exception of grazing angles $\theta $ around $\left| \theta \right|
\lessapprox \frac{\pi }{2}$. In Fig.(1) we plot the radiated intensity $I(%
\overline{\omega })=\int_{-\frac{\pi }{2}}^{\frac{\pi }{2}}\frac{d\theta }{%
\pi }I\left( \theta ,\overline{\omega }\right) $ vs. scaled frequency $%
\overline{\omega }$ for various vortex speeds. Values for the gap, the
quasiparticle damping $\gamma $ and the vortex speeds are taken as indicated
in the Figure caption. The radiated intensity $I(\overline{\omega })$
displays a sharp peak at a frequency 
\begin{equation}
2\pi \nu _{peak}=\omega _{peak}=\frac{1}{\sqrt{3}}\frac{\left| v_{x}\right| 
}{\lambda \left( \gamma \right) }
\end{equation}
This is in agreement with the physical picture, that it takes a time $%
T\simeq \frac{2\lambda _{L}}{\left| v_{x}\right| }$ for the London core of a
vortex to pass the interfacial region. Our analytical estimatation for $%
\omega _{peak}$ applies provided the frequency of the emitted radiation
fulfills $\hbar \omega <2\Delta $. Indeed, even for a very high vortex speed
around $\left| v_{x}\right| \simeq v_{F}$ \ on finds $\hbar \omega _{peak}=%
\frac{1}{\sqrt{3}}\frac{\hbar v_{F}}{\lambda \left( \gamma \right) }\simeq 
\frac{\xi _{BCS}}{\lambda _{L}}\cdot \Delta \ll \Delta $ , since $\xi
_{BCS}\simeq \frac{\hbar v_{F}}{\Delta }\ll \lambda _{L}$. Rather high
vortex speeds around $v_{x}\simeq 10^{3}\frac{m}{\sec }$ have been measured
in layered type-II materials in the experiments of Doettinger et al. \cite
{Doettinger}. For a London penetration depth of order $\lambda \left( \gamma
\right) \simeq 10^{3}$\AA\ \ we expect the position of the peak at $\nu
_{peak}=\frac{\omega _{peak}}{2\pi }\simeq \frac{1}{\sqrt{3}2\pi }\frac{%
10^{3}\frac{m}{\sec }}{10^{3}\text{\AA }}\simeq 1\,GHz$ , a typical
microwave frequency. Also the height of the peak intensity increases
linearly with the vortex speed $v_{x}$, as shown in Fig.(1), in agreement
with our analytical estimate $I\left( \overline{\omega }_{peak}\right)
\simeq \allowbreak \frac{3}{4}\sqrt{3}\,\frac{\overline{v}_{x}}{\overline{%
\lambda }}$. \ Finally, the total amount of energy that gets emitted into
free space due to the transition radiation effect follows integrating over
all frequencies. Certainly, at very high frequencies, say for $\omega
>\omega _{pl}$ , our local electrodynamics becomes invalid. But since the
integral converges rapidly the error is small if we simply extend the upper
integration limit towards infinity. We obtain: 
\begin{equation}
\frac{\varepsilon _{rad}}{L}\simeq \ \left[ \frac{\Phi _{0}}{2\pi \lambda
\left( \gamma \right) }\right] ^{2}\ \left( \frac{v_{x}}{c}\right) ^{2}
\end{equation}
Assuming $\lambda _{L}\simeq 100\,nm$\ and a vortex speed of order $\frac{%
v_{x}}{c}\simeq 10^{-5}$ gives $\varepsilon _{rad}\simeq \ 10^{-22}\,W\cdot
sec$ (per length $L=1\,cm$). Due to the large impedance mismatch of the
superconductor with the vacuum ($Z_{0}\simeq 377\,\Omega $) it is not
surprising that only a tiny fraction of the nucleation energy\cite{Abrikosov}%
,\cite{de Gennes} of a single Abrikosov vortex gets radiated into free
space. So, when the line is destroyed during its passage through the
interfacial region, the energy that is not radiated into free space gets
converted into elementary excitations (e.g. polaritons) and, eventually,
into heat.

A larger effect arises when several vortices, say a bundle of altogether $%
\left| \Lambda \right| $ flux lines, move coherently at constant speed ${\bf %
v}$ towards the surface of a superconductor. According to the superposition
principle the (effective) drive term for several flux lines, situated at
time $t=0$ at positions ${\bf r}_{L}=(x_{L},y_{L})$ inside the
superconductor, follows from Eq.(\ref{self fields}) upon summation over the
instantaneous positions ${\bf R}_{L}=$ ${\bf r}_{L}+{\bf v\ }t$ of the
individual vortices (and the instantaneous positions of the mirror vortices
at $\widetilde{{\bf R}}_{L}=\widetilde{{\bf r}}_{L}+\widetilde{{\bf v}}{\bf %
\ }t$). The linearity of the dynamical London equations Eq.(\ref{dynamical
London equations}) implies that the electromagnetic field ${\bf E}^{\left(
\Lambda \right) }$ generated by such a bundle of flux lines is: 
\begin{equation}
{\bf E}^{\left( \Lambda \right) }{\bf (}x,k_{y}{\bf ,}\omega {\bf )}=\sum_{%
{\bf r}_{L}\epsilon \Lambda }e^{-i\left( k_{y}y_{L}+K_{\omega }x_{L}\right) }%
{\bf E(}x,k_{y}{\bf ,}\omega {\bf )}
\end{equation}
From the theory of diffraction of electromagnetic waves it follows that the
associated Poynting flux, and hence the radiated peak intensity emitted by a 
{\em periodic} arrangement of coherently moving flux lines (for example, a
square vortex lattice $\Lambda $, with lattice constant $d_{\Lambda }$,
consisting out of $\left| \Lambda \right| =N_{x}\cdot N_{y}$ flux lines),
should be increased by a factor proportional to the number of participating
modes compared to the case where only unlocked modes are superposed. The
total radiated energy (per unit length $L$ of the vortex lines) is\ 
\[
\frac{\varepsilon _{rad}^{\left( \Lambda \right) }}{L}=\int_{-\infty
}^{\infty }\frac{d\omega }{2\pi }\frac{c}{4\pi }\int_{-\left| \frac{\omega }{%
c}\right| }^{\left| \frac{\omega }{c}\right| }\frac{dk_{y}}{2\pi }\frac{%
\left| \frac{\omega }{c}\right| \,\left| a_{V}\left( k_{y}{\bf ,}\omega
\right) \right| ^{2}\,\left| S_{\Lambda }(k_{y},\omega )\right| ^{2}}{\sqrt{%
\frac{\omega ^{2}}{c^{2}}-k_{y}^{2}}}\, 
\]
The structure factor of the moving vortex lattice 
\begin{equation}
S_{\Lambda }(k_{y},\omega )=\left| \sum_{{\bf r}_{L}\epsilon \Lambda
}e^{-i\left( k_{y}\,y_{L}+K_{\omega }x_{L}\right) }\right| ^{2}
\end{equation}
represents, for large $N_{x}$ and $N_{y}$, a series of delta spikes,
selecting wave numbers $k_{y}$ and frequencies $\omega $ such that $k_{y}=%
\frac{2\pi }{d_{\Lambda }}n_{y}$ and $\omega =\frac{2\pi }{d_{\Lambda }}%
\left( n_{x}v_{x}+n_{y}v_{y}\right) $, with $n_{x}$ and $n_{y}$ integer. For
intermediate magnetic fields $B_{c1}\ll B\ll B_{c2}$ the lattice constant $%
d_{\Lambda }$ is much larger than the coherence length $\xi $, but smaller
than the London depth $\lambda _{L}$ , so that $\xi \ll d_{\Lambda }\ll
\lambda _{L}$. We derive from this a simple estimation for the radiation
energy emitted into free space by alltogether $\left| \Lambda \right| \gg 1$
coherently moving vortex lines (arranged in the geometry of a square lattice
as described above): 
\begin{equation}
\frac{\varepsilon _{rad}^{\left( \Lambda \right) }}{L}\simeq \frac{\left|
\Lambda \right| }{\pi }\,\left( \frac{\Phi _{0}}{d_{\Lambda }}\right) ^{2}\,%
\frac{\left| v_{x}\right| }{c}
\end{equation}

The principal result is that the transition radiation effect for a cluster
of altogether $\left| \Lambda \right| \gg 1$ {\em coherently} moving flux
lines is a {\em linear} function of the vortex speed $\left| v_{x}\right| $.
The emitted power {\em per vortex} is huge compared to the radiation emitted
by a single vortex by a factor proportional to the number of vortices, since 
$\frac{1}{d_{\Lambda }^{2}}\propto \left| \Lambda \right| $. This
enhancement is due to {\em constructive} interference, like in the super
radiation effect of laser physics. Assuming $\lambda _{L}\simeq 100\,nm$\
and $\ d_{\Lambda }=0.5\,\lambda _{L}$ gives a total of $\left| \Lambda
\right| \simeq 0.25\times 10^{12}$ vortices per $cm^{2}$. This leads (per
length $L=1\,cm$) and a vortex speed $v_{x}\simeq 10^{-5}c$ , $v_{y}=0$ to a
value of order $\varepsilon _{rad}^{\left( \Lambda \right) }\simeq
0.12\times 10^{-3}\,W\,sec$. This estimation suggests that a sensitive S-I-S
Josephson radiometer could be used to detect the predicted spectrum of
transition radiation in free space. The sharp peak of the radiated intensity
around $\omega \sim \frac{\left| v_{x}\right| }{\lambda }$ allows to measure
MPD for a known vortex speed, and vice versa.

Our calculations indicate that transition radiation is a {\em generic}
phenomenon, that should exist in other types of {\em ordered} condensed
matter containing mobile (topological) defects. We expect transition
radiation to be emitted into free space by rapidly moving dislocations, by
magnetic bubbles and Bloch walls in magnets, and also by travelling polarons.

{\bf Acknowledgments}: We thank V.L. Ginzburg and A.A. Abrikosov for the
interest they have shown into our work and for correspondence. It is a
pleasure to thank D. Rainer for an illuminating discussion of the thin
solenoid model, and to thank R.P. Huebener for useful information on high
speed flux-flow experiments.



\begin{figure}[tbp]
\caption{Frequency distribution $I(\overline{\protect\omega })$ of intensity
of transition radiation emitted into free space for different vortex speeds;
the labels $a), b), c), d)$ correspond to $v_y=0$ and to $v_{x}/c=2\times
10^{-5}, 1.5\times 10^{-5}, 10^{-5}, 5\times 10^{-6}$, respectively. Assumed
values for scattering rate $\protect\gamma/\protect\omega_{pl}=10^{-4}$ and
gap $\Delta/\protect\omega_{pl}=10^{-3}$. }
\label{Fig.1}
\end{figure}

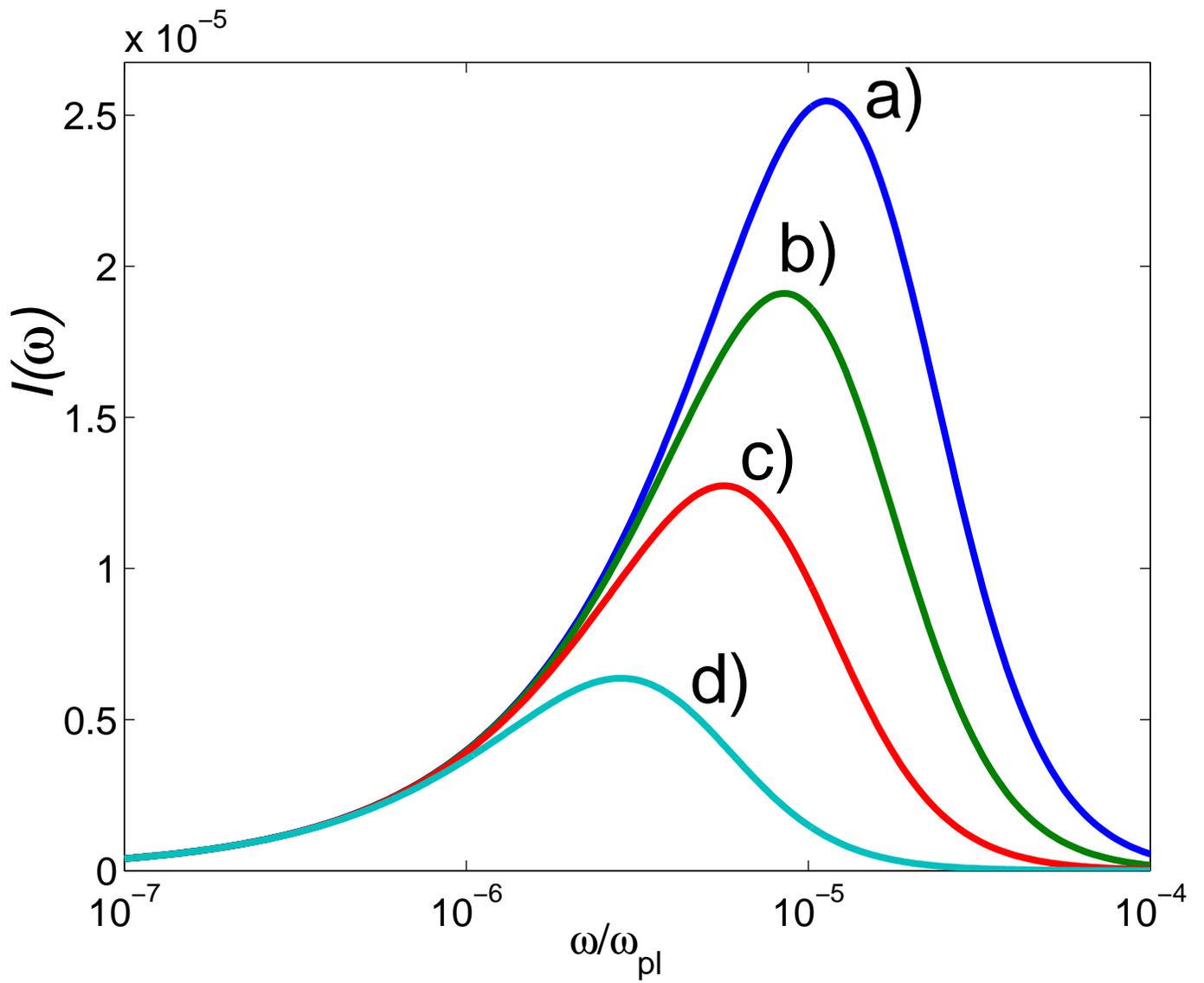
\end{document}